\documentclass[12pt]{article}
\usepackage{pdproc,epsfig,deluxe}

\begin{document}

\renewcommand{\thefootnote}{\alph{footnote}}

\title{THE ORIGIN OF COSMIC RAYS AT ALL ENERGIES}

\author{ARNON DAR}

\address{ Department of Physics and Space Research 
Institute,\\Technion, Haifa 32000, Israel\\
{\rm E-mail: arnon@physics.technion.ac.il}}

\abstract
{There is mounting evidence from observations of long duration gamma ray 
bursts (GRBs), supernova remnants (SNR) and the supernova (SN) explosion 
1987A, that SN explosions eject highly relativistic bipolar jets of 
plasmoids (cannonballs) of ordinary matter. The highly relativistic 
plasmoids sweep up the ambient matter in front of them, accelerate it to 
cosmic ray (CR) energies and disperse it along their long trajectories in 
the interstellar medium, galactic halo and intergalactic space. Here we 
use the remarkably successful cannonball (CB) model of GRBs to show that 
bipolar jets from Galactic SN explosions can produce the bulk of the 
Galactic cosmic rays at energies below the ankle, while the CRs which 
escape into the intergalactic space or are deposited there directly by 
jets from SNe in external galaxies can produce the observed cosmic ray 
flux with energies above the ankle. The model predict well all the 
observed properties of cosmic rays: their intensity, their spectrum 
including their elemental knees and ankles, their composition and the 
distribution of their arrival directions.  At energies above the CR ankle, 
the Galactic magnetic fields can no longer delay the free escape of such 
ultra high energy CRs (UHECRs) from the Galaxy. These UHECRs, that are 
injected into the intergalactic medium (IGM) by the SN jets from our 
Galaxy and all the other galaxies and are isotropized there by the IGM 
magnetic fields, dominate the flux of UHECRs. Almost all the extragalactic 
UHECRs heavier than helium photo-disintegrate in collisions with the far 
infrared (FIR), microwave and radio background radiations. The surviving 
CR protons and He nuclei suffer a Greisen-Zatsepin-Kuzmin (GZK) cutoff due 
to pion photo-production in collisions with the FIR, microwave and radio 
background photons.}

\normalsize\baselineskip=15pt
 
\section{Introduction}
\noindent
Cosmic rays (CRs) were discovered by Victor Hess in 1912. Today, 93 years
later, their origin is still debated.  CRs have been studied in
experiments above the atmosphere, in the atmosphere, on the ground,
underground and in space. Their energies cover an enormous range, from sub
GeV to more than a few $10^{11}\, GeV\,,$ over which their differential
flux decreases by roughly 33 orders of magnitude. Any successful theory of
the origin of CRs must explain their main observed properties near Earth
(for recent reviews see Biermann \& Sigel 2002; Olinto 2004; Cronin 2004
Watson 2004; Hoerandel 2005; Engel 2005): 

\noindent
{\bf The CR energy spectrum} shown in Fig. 1, has
been measured up to hundreds $EeV$. It can be
approximated by a broken power-law, $dn/dE\sim E^{-p}\,,$ with a series of
breaks near $ 3\, PeV$ known as `the knee', a second `knee' near
200 PeV and an `ankle' near $3\, EeV\, $. The power-law index
changes from $p\sim 2.67\, $ below the knee to $p\sim 3.05$ above it and
steepens to $p\sim 3.2$ at the second knee. At the ankle the
spectral index seems to decrease to $p\sim 2.7\, .$ The spectral behaviour
above $50\ EeV$ is still debated (see e.g., Olinto 2004; Cronin 2004;
Watson 2004).

\noindent 
{\bf The CR elemental composition} is known well only from direct
measurements on board satellites which run out of statistics well below
the knee energy. The measured composition of high energy CRs is highly
enriched in elements heavier than hydrogen relative to that of the solar
system.  The enrichment increases with
atomic number and with CR energy almost up to  the second knee beyond
which it appears to decline (e.g., Kampert et al.~2004; Hoerandel, 2004 ).
as shown in Figs 4,5.
It is still debated at energies above the ankle (e.g. Cronin 2004; 
Watson 2004).  The
detailed elemental composition at the knee and above it is known only very
roughly (e.g. Hoerandel 2004, 2005; Cronin 2004; Watson 2004; Engel 2005).

\noindent
{\bf The CR arrival directions} at energy below $\sim EeV$ are 
isotropized by the Galactic magnetic fields. Only at energies
well above $EeV\,,$ their arrival directions 
may point towards their Galactic sources and only at extremely high 
energies they may point towards nearby extragalactic sources.  
Initial reports of deviations from isotropy in the arrival directions 
of such cosmic rays with energy above $EeV$ and of some clustering in their 
arrival directions are still debated (e.g. Cronin 2004).

\noindent
{\bf The Galactic cosmic ray luminosity} has been estimated to be 
between $10^{41}\, erg\, s^{-1}$ (e.g., ) and more than
$10^{42}\, erg\, s^{-1}\, $ (see, e.g. Dar \& De R\'ujula~2001)
depending on the size of the Galactic CR halo.

It is widely believed that Galactic CRs with energy below the knee are
accelerated mainly in Galactic supernova remnants (SNRs). The opinions on
the origin of CRs with energy between the knee and the ankle are still
divided between Galactic and extragalactic origin.  CRs with energy above
the ankle are generally believed to be extragalactic in origin because
they can no longer be isotropized by the Galactic magnetic fields while
their observed arrival directions are isotropic to a fair approximation. 
Yet Galactic origin is not ruled out -- they may be produced by the decay 
of unknown massive particles or other sources which are 
distributed
isotropically in an extended Galactic halo (e.g. Plaga 2002).
An observational proof that
the CRs above the ankle are extragalactic in origin, such as arrival
directions which are correlated with identified extragalactic sources or a
Greisen-Zatsepin-Kuzmin (GZK) cutoff due to $\pi$ production in their
collisions with the cosmic microwave background radiation (MBR), are still
lacking (Cronin 2004). 
Moreover, there is no single solid observational evidence which supports
the SNR origin of the bulk of CRs below the knee (see, e.g. Plaga 2002 and 
references
therein).  In fact, the evidence from $\gamma$-ray astronomy, x-ray
astronomy and radio astronomy strongly suggest that {\bf SNRs are not the
major accelerators of CRs with energy below the knee}: 

\begin{itemize} 
\item{}
{\bf SNR origin cannot explain the Galactic CR luminosity:}
Radio emission and X-ray emission from SNRs provide strong evidence 
for acceleration of high energy electrons in SNRs. Some SNRs were
also detected in TeV $\gamma$-rays which could be produced by the prompt 
decay of $\pi^0$'s from collisions of high energy hadronic 
cosmic rays with the ambient protons and nuclei in and around the SNR. 
However, recent observations of SNRs inside molecular clouds 
(and careful analysis of Tev emission in others, such as SN1006) 
show that the time-integrated CR luminosity of these SNRs does not   
exceed a few $10^{48}\, erg$ unless an extremely small mean baryon density 
is assumed for the production region.
The rate of Galactic SN explosions, which
is estimated to be $\sim 1/50\, year\, ,$ yields a total CR
luminosity which falls short by  two orders of magnitude  than the 
estimated luminosity of the Milky Way (MW) in CRs, $L_{CR}[MW] > 10^{41}\, 
erg\, s^{-1}\, .$
\item{} 
{\bf SNR origin cannot explain the diffuse Galactic GBR:} 
The
interactions of CR electrons with ambient photons and of CR nuclei with
ambient nuclei in the interstellar medium produce a diffuse background of
$\gamma$-rays. Such a diffuse gamma background radiation (GBR) has been
detected by EGRET and Comptel on board the Compton Gamma Ray Observatory
(CGRO). However, the scale length of the distribution of SNR in the
Galactic disk is $\sim 4\, kpc$ and cannot explain the scale length of the
observed GBR, which is larger by more than an order of magnitude. In
particular, energetic electrons cool rapidly by inverse Compton scattering
of stellar light and of microwave background photons. Over their
cooling time they cannot reach by diffusion far enough from the SNRs to
explain the intensity of GBR at large Galactic distances. 

\item{}
{\bf SNR origin cannot explain the diffuse radio emission from galaxies
and clusters:} Because of their fast cooling in the MBR the electrons from
SNRs, which are mainly located in the galactic disk, cannot reach large
distances from the galactic disk by diffusion or galactic
winds. The radio emission from our Galaxy, from edge-on galaxies and from
the intergalactic space in clusters of galaxies provide evidence for high
energy electrons at very large distances from the galactic disks where
most of the SNRs are located. 
\end{itemize}

\noindent 

Although supernova remnants (SNRs) do not seem to be the main source
of CRs in our Galaxy, in external galaxies and in the intergalactic medium
(IGM), still SN explosions may be the main source of CRs at all energies
if SNe emit highly relativistic bi-polar jets which produce the visible
GRBs when they point in our directions (Dar and Plaga 1999)\footnote{The
association between GRBs and high energy CRs was first suggested by Dar et
al. (1992). Waxman (1995), Vietri (1995) and Milgrom (1995) 
suggested that extragalactic GRBs are the main accelerators only of  
ultra-high energy 
cosmic rays (UHECRs). However, following the first
observational evidence for a GRB-SN association (Galama et al 1998),
Dar \& Plaga (1999), 
suggested that the bulk of the cosmic rays {\bf at all energies} are
accelerated in bipolar jets that are ejected in Galactic SN explosions 
which produce Galactic GRBs, most of which are beamed away from Earth.}. 
Here, I will
outline briefly a simple theory of the origin of CRs at all energies
which is based on the extremely successful cannonball (CB) model of GRBs
(Dar \& De R\'ujula~2000,2004). I will show that it explains
remarkably well the main observed properties of Galactic and extragalactic
CRs.  The complete theory and a rigorous derivation of the main
properties of CRs from the CB model will be published elsewhere  (Dar \& 
De R\'ujula~2005b). According to this theory: 

\begin{itemize} 
\item{} 
The bulk of the CR nuclei are accelerated by the highly relativistic
bipolar jets of plasmoids (cannonballs) of ordinary matter ejected in SN
explosions (which produce GRBs, most of which do not point towards Earth).
These jets can accelerate swept in ISM particles to the highest observed
cosmic ray energies\footnote{Jets from microquasars and active galactic
nuclei may also contribute to CR acceleration in galaxies
and in clusters of galaxies, but they will not be discussed here.}.

\item{}
The elemental knees  are the maximal lab
energy that ISM nuclei acquire by elastic magnetic scattering in the
CBs' rest frame.  The knee in the al-particle spectrum is the maximal lab
energy of CR protons and the `second' knee is that of the iron group
nuclei and heavier metals.

\item{} 
The Galactic cosmic rays escape outside the Galaxy by diffusion in the
Galactic magnetic fields. Their energy-dependent confinement time steepens
their Galactic spectrum. The ankle is the energy where the Galactic
magnetic fields can no longer isotropize their directions and delay 
their free escape into the IGM.

\item{} 
The CRs from SN explosions that have escaped the galaxies into the
IGM, or were deposited there directly by the jets from galactic SNe, 
produce the bulk of the CRs in the intra cluster medium in galaxy clusters 
and in
the IGM outside clusters.  They have been accumulating there over the
Hubble time and were isotropized by the IGM magnetic fields. 
Above the ankle, the Galactic magnetic fields cannot shield it from 
UHECRs. Thus most of he
UHECRs nuclei in the Galaxy with energy above the ankle are extragalactic
in origin.

\item{}
The galactic CRs which escape into the IGM, or are directly deposited 
there, have the 
injection spectral
index $p\approx 2.17$ below the elemental knees. 

\item{}
The spectrum of the extragalactic UHECRs in the IGM
is modified by the cosmic expansion
and by their interaction 
with the far infrared, microwave and radio
background radiations. The effective thresholds for  
pair production and photo-disintegration are well below the CR 
ankle but the Galactic magnetic field and Galactic winds 
probably prevent their penetration at energies well
below the ankle. At energies above the 
ankle, the UHECRs which cross through the Galaxy
dominate the CR flux. Most of the UHECR nuclei
heavier than $He$ are destroyed by photo-disintegration in less than a
Hubble time. The bulk of the UHECR protons and $He$ nuclei in the IGM 
suffer a GZK suppression due to $\pi$ production and only those from 
nearby extragalactic sources may reach the Galaxy.

\item
Direct CR deposition in the IGM, by CBs and their associated CR jets
and by CR diffusion and galactic winds, stir it 
up and generate the turbulent IGM magnetic fields (Dar and De R\'ujula 
2005a).

\item{}
The CR electrons, in external galaxies and in the IGM  
produce the bulk of the 
extragalactic diffuse gamma-ray background radiation (GBR) by
inverse Compton scattering of the cosmic microwave background (MBR) (Dar
and De R\'ujula 2000).
\end{itemize}

\noindent
\section{Collimated jets and relativistic beaming} 
\noindent
Radio, optical and X-ray observations with high spatial
resolution indicate that relativistic jets, which are fired by quasar and
microquasars, are made of a sequence of plasmoids (cannonballs) of
ordinary matter whose initial expansion (presumably with an expansion
velocity similar to the speed of sound in a relativistic gas) stops
shortly after launch (e.g., Dar \& D R\'ujula~2004 and references
therein). The turbulent magnetic fields in such plasmoids gather and
scatter the ionized ISM particles on their path. For the sake of
simplicity, we shall assume here
that all the incident ISM
particles which are scattered in the CBs 
are first isotropized there by magnetic scattering.
Electrons which are trapped in the CBs cool there
quickly by synchrotron emission.  This radiation which is emitted
isotropically in the CBs' rest frame is beamed by the relativistic bulk
motion of the CBs (Lorentz factor $\gamma=1/ \sqrt{1-\beta^2}\, )$. Let
primed quantities denote their values in the plasmoid's rest frame and
unprimed quantities their corresponding values in the lab frame. Then the
angle $\theta'$ of the emitted photons in the CBs' rest frame relative to
the CBs' direction of motion, and the corresponding angle $\theta$ in the
lab frame, are related through: 
\begin{equation}
  \cos\theta' = {\cos\theta-\beta \over 1-\beta\, \cos\theta}\, .
\label{thetaprime}
\end{equation}
This relation is valid to a good approximation also for the emission of 
highly relativistic massive particles. When applied to 
an isotropic distribution of emitted particles 
in the CBs' rest frame, it yields a distribution, 
\begin{equation}
{dn\over d\Omega}={dn\over d\Omega'}\, {dcos\theta'\over dcos\theta}
\approx {n\over 4\, \pi}\, \delta^2
\label{dist}
\end{equation}
in the lab frame where
\begin{equation}
  \delta = {1\over \gamma\, (1-\beta\, \cos\theta)}
\label{Doppler}
\end{equation}
is the Doppler factor of the CB motion viewed from a lab angle $\theta$. 
For plasmoids with highly relativistic bulk motion Lorentz factor, 
$\gamma^2 \gg 1\, ,$  and for $\theta^2 \ll 1$, 
the Doppler factor is well approximated by 
\begin{equation}
\delta \approx {2\gamma\over 1+\gamma^2\, \theta^2}\; .
\label{Dopplerapprox}
\end{equation}
Hence, the isotropic distribution of the emitted particles in the CB's rest 
frame is collimated into a narrow conical beam, ``the beaming cone'', 
around  the direction of motion of the CB in the lab frame,
\begin{equation}
{dn\over d\Omega}\approx {n\over 4\, \pi}\, \left[ {2\gamma\over 
1+\gamma^2\, \theta^2}\right]^2\, .
\label{distapprox}
\end{equation}
The beaming depends only on the CB's Lorentz factor but not on the mass of 
the scattered particles. 

\noindent
Ambient ISM particles 
which are practically at rest in the ISM, enter the CBs with an energy
$E'=\gamma\, m\, .$ After magnetic scattering and isotropization in the 
CB they are emitted with a lab energy 
\begin{equation}
E=\gamma\, E'\, (1+\beta^2\, cos\theta')\, .
\label{CRE}
\end{equation}
Their energy distribution in the lab frame given by a simple step function:  
\begin{equation}
{dn\over dE}= 
{dn\over dcos\theta'}\, {d\cos\theta'\over dE}\approx {n\over 2\, 
\beta^2\,\gamma^2\, m}\,\Theta(E-(2\,\gamma^2-1)\, m)\, , 
\label{Edist}
\end{equation}
where $\Theta(x)=1$ for $x<1$ and  $\Theta(x)=0$ for $x>1\, .$  
The accelerated nuclei ($m=A\, m_p$) and electrons ($m=m_e$), which are
initially ejected into a narrow cone, are later scattered and isotropized 
by the
galactic magnetic fields. The electrons do not reach far away 
from their 
injection
cone before they radiate most of their initial energy via synchrotron
emission and inverse Compton scattering of the microwave background
photons  along their direction of motion. Their radiation, which is beamed
along their motion, can be seen by an observer outside the CBs' beaming
cone, only when their direction of motion points towards him. 
Because of the fast radiative cooling of energetic electrons, their
radiation is visible mainly when they are still within/near the beaming
cone. The gradual increase of the opening angle of the injection cone due
to jet deceleration, together with the finite lifetime of the radiating
electrons which confines them to near the jet, produce the
a conical images of radio and optical jets. It is often confused
with the true geometry of the relativistic jet, which 
reveals itself only in observations at much higher frequencies 
where the observed emission requires much 
stronger magnetic fields than those present in the ISM and IGM.

\section{GRBs from SN jets}
\noindent
There is mounting observational evidence that long duration gamma ray
bursts (GRBs) are produced by ultra-relativistic jets of ordinary matter
which are ejected in core collapse supernova (SN) explosions (see, e.g.
Dar 2004; Dado these proceedings) as long advocated by the remarkably
successful CB model of GRBs (e.g., Dar \& De R\'ujula~2000, 2004; Dado et
al.~2002, 2004 and references therein). In the CB model,
the long-duration GRBs are produced in ordinary core-collapse SN
explosions. Following the
collapse of the stellar core into a neutron star or a black hole, and
given the characteristically large specific angular momentum of stars, an
accretion disk or torus is hypothesized to be produced around the newly
formed compact object, either by stellar material originally close to the
surface of the imploding core and left behind by the explosion-generating
outgoing shock, or by more distant stellar matter falling back after its
passage (De R\'ujula 1987). A CB is emitted, as observed in microquasars,
when part of the accretion disk falls abruptly onto the compact object
(e.g.~Mirabel \& Rodrigez 1999; Rodriguez \& Mirabel 1999 and references
therein). The high-energy photons of a single pulse in a GRB
are produced as a CB coasts
through the ``ambient light'' permeating the surroundings of the
parent SN. The electrons enclosed in the CB Compton
up-scatter photons to energies that, close to the CBs direction
of motion, correspond to the $\gamma$-rays of a GRB and
less close to it, to the X-rays of an XRF.
Each pulse of a GRB  corresponds to one
CB in the jet. The timing sequence of emission of the successive 
individual
pulses (or CBs) reflects the chaotic accretion process and its
properties are not predictable, but those of the single pulses are (Dar \&
De R\'ujula~2004 and references therein).

\section{CR acceleration by decelerating SN jets}
\noindent
Let $n_{_A}$ be the density of nuclei of atomic mass $A$ along the 
trajectories of SN jets. Let us assume that the elemental abundances
$X_A=n_{_A}/n_b\,,$ are constant along the trajectories, 
where $n_b$ is the total baryon density. 
Let us define a total effective mass of these particles, \,$\bar{m}=m_p\,
\Sigma A\, X_A\, .$ Let us assume that most of them are swept into the jet. 
Let us also neglect the small energy radiated away by the swept up
electrons except for the fact that it ionizes completely the ISM in front
of the jet. Then, for large values of $\gamma$, energy conservation 
implies that 
$M(x)\, \gamma(x)=M_0\,\gamma_0$ where $M(x)$ and $\gamma(x)$ are the 
total rest-mass and
bulk-motion Lorentz factor of the jet along its trajectory with initial
values, $M_0=M(0)$ and $\gamma_0=\gamma(0)\, ,$ respectively. Let us
denote by $dN_{_A}(x)$ the number of cosmic ray nuclei of atomic mass $A$
which enter a CB at a distance $x\, .$ Consequently 
one can relate the CB deceleration to the number of ISM 
particles which enter the CB with a Lorentz factor $\gamma\, .$ 
\begin{equation} M\, d\gamma = - dN_b\, \bar{m}\, \gamma^2\,, 
\label{dec1} 
\end{equation} 
or 
\begin{equation}
dN_A\approx {X_A\, M_0\, \gamma_0 \over \bar{m}}\, {d\gamma\over
\gamma^3}\, . 
 \label{dec2} 
\end{equation} 
Note that Eq.~(\ref{dec1}) depends neither on the geometry of the jet 
nor on the density profile along the jet trajectory.
If a constant fraction of these nuclei are scattered elastically 
by the CBs then
the CR energy spectrum due to elastic scattering 
by a decelerating jet can be obtained by 
replacing $n_p$ in Eq.~(\ref{Edist}) by $dn_{_A}$ and integrating over 
$\gamma$ at a fixed $E$ under condition~(\ref{CRE}): 
\begin{equation} 
{dN_A\over dE}=
          {X_A\, M_0\, \gamma_0 \over 2\, A\, m_p\, \bar{m} }\,
   \int {d\gamma\over \gamma^5}\propto {X_A\over A\, \bar{m}}\,
 \left[\left[{E\over A\, m_p}\right]^{-2}-
 \left[{E_{max}\over A\, m_p}\right]^{-2}\right]\,\Theta(E-E_{max}), 
\label{CRSbk}
\end{equation}
where, 
\begin{equation}
E_{max}=2\, A\, m_p\,(\gamma_0^2-1) 
\label{Emax}
\end{equation}
is the maximum energy of elastically scattered 
ISM nuclei in the CB rest frame. 
For energies well below $E_{max}$, the injected spectrum of 
CRs is a simple power-law, $dN_A/dE\sim E^{-2}$.   
The effective power on the rhs of Eq.~(\ref{dec1}) and consequently on 
the rhs of Eq.~(\ref{CRSbk}) steepens slightly
if the emission of the elastically scattered ISM particles from the CBs 
is delayed by diffusion in the CB's magnetic field
whose radius increases during its deceleration.
In the CB model, the CB's radius increases like $\gamma^{-2/3}$ (Dado et 
al.~2002) which yields $dN_A/dE\sim E^{-13/6}\approx E^{-2.17}$ 
(Dar \& De R\'ujula~2005b).
Such a CR injection spectrum was deduced by us from the Galactic
diffuse gamma ray emission by inverse Compton scattering of light
and microwave background radiation   
by cosmic ray electrons (e.g. Dar and
De R\'ujula 2001) and from the diffuse radio emission  
of galaxies and galaxy clusters by synchrotron radiation of CR
electrons in their magnetic fields. 
A similar power-law index has been claimed to arise in numerical 
calculations
of CR acceleration in collisionless shock acceleration (Bednarz \&
Ostrowski 1998; Kirk et al.~2000).  

\section{Spectral steepening by magnetic trapping} 
\noindent
The cosmic rays which are accelerated by the highly relativistic jets from
SN explosions are initially beamed into narrow cones along the jets'
trajectories. Their free escape into the intergalactic space is delayed by
diffusion in the galactic magnetic fields which isotropize their direction
of motion. 
The energy-dependent diffusion of CRs due to pitch-angle scattering with a
Kraichnan spectrum of MHD turbulence results in a CR
residence time in the host galaxy which behaves like (e.g. Wick, Dermer
\& Atoyan 2003) $(E/Z)^{-c}$ with $c\approx 0.5$ to be compared with
$c=0.6\pm 0.10$, which was inferred from the observed relative abundances 
of CRs  
at energies far below the knee (Swordy et al.~1990). The same diffusion 
law inside the CB enhances the emission of the swept-in nuclei by a 
factor $(A/Z)^c.$ Thus, the diffusion of CRs outside the CBs
and  the accumulation of CRs before
their escape from the Galaxy steepens their injection spectral index  
$2.17$ to $p=2.17+c=2.67\pm 0.10$, which yields,
\begin{equation}
{dN_A\over dE} \propto {X_A(E/A)\, A^{p-1}\ \over \bar{m}}\, 
                      \left[{E\over m_p}\right]^{-c}\, 
                      \left[\left[{E\over  m_p}\right]^{-2.17}-
                      \left[{E_{max}\over m_p}\right]^{-2.17}\right]\,
                      \Theta(E-E_{max})\, . 
\label{CRdnde} 
\end{equation} 
The decreasing metalicity and the  deceleration of the jet 
along the jet's trajectory  result in  an effective energy dependent  
$X_A(E/A)$ which is discussed in section 7. 
\section{The CR knees}
\noindent
The distribution of the initial Lorentz factor,
$\gamma_0\,$ of the CBs in the bipolar SN jets which produce GRBs
was deduced from a cannonball model analysis of the afterglow 
of GRBs with well known redshift (Dar \& De 
R\'ujula~2004). Their well fitted log-normal distribution 
around $\gamma_0\approx 1250$ yields
a log-normal distribution of $E_{max}$ around 
$\bar{E}_{max}\approx 3\,A\, PeV\,,$ which can be well approximated 
by :
\begin{equation}
P(E_{max})\approx  {1\over E_{max}\,\sigma\, \sqrt{2\, \pi}}\, 
   exp\left[-{(lnE_{max}-ln\bar{E}_{max})^2\over 2\, \sigma^2}\right]\, . 
\label{PEmax}
\end{equation}
The energy-spectra of the individual CR elements
is obtained by integrating Eq.~(\ref{CRdnde}) over $E_{max}$ with the 
probability distribution~(\ref{PEmax}):
\begin{equation}
{dN_A\over dE}\rightarrow \int P(E_{max})\, {dN_A\over dE}\, dE_{max}\, . 
\label{CREspectrum}
\end{equation}
The narrow log-normal distribution of the Lorentz factors of CBs, as 
inferred from GRBs (Dar \& De
R\'ujula~2004), produces a narrow distribution of the maximum energy of 
CRs produced by elastic scattering of ambient ISM particles. 
Thus, the energy spectra of the individual CR nuclei, integrated over all 
CBs, retain a sharp knee at, 
\begin{equation}
         E_{knee} \approx \bar{E}_{max}= 2\, A\, m_p\, 
(<\gamma_0^2-1>) 
                    \approx 3\,A\, PeV\, .  
\label{Eknee} 
\end {equation}
Fig.~(\ref{fig4}) compares the predicted energy spectra of H, He,
and Fe group nuclei around their respective knees as
calculated\footnote{For the sake of simplicity, I have neglected here
Fermi acceleration inside the CBs and I have used a fitted effective
$\sigma$.} from Eqs.~(\ref{CRdnde}), (\ref{CREspectrum}), (\ref{Eknee}),
and those extracted from the KASKADE observations (Kampert et al. 2004;
Hoerandel et al 2004).  Despite the large experimental uncertainties, the
theory seems to reproduce correctly the elemental knees, their
A-dependence and their energy spectrum around these knees. Note that the
predicted knee for different elements is proportional to $A$ rather than
to $Z$ as in conventional models based on shock acceleration, where the
maximum energy gain is limited by the requirement that the size of the
accelerator be larger than the Larmor radius of the accelerated CR
particles. This difference is large only for nuclei lighter than He
nuclei.  However, accurate values of the knee energy of such 
nuclei are not available yet from the CR experiments.

\noindent
The energy spectrum of individual elements falls rapidly at their knees.
The all-particle CR spectrum beyond the proton knee loses progressively
the contribution from heavier and heavier elements, until it becomes
almost pure iron.  This produces the steepening of the spectrum between
the proton knee and the iron knee and results in a composition which
gradually approaches a pure iron+heavier metal composition. Because the
abundances of elements heavier than iron are rather small compared to
iron, {\bf the all-particle spectrum steepens at the iron knee forming the
second knee} in the all-particle spectrum. 

\section{Elemental abundances and spectral indexes below the knees}
\noindent
The elemental abundances of CRs accelerated 
by SN jets can be read from Eq.~(\ref{CRdnde}),
\begin{equation}
X_A[CR] = X_A(E/A)\, A^{p-1}\, ,
\label{CRabund}
\end{equation}
where $X_A(E/A)$ is the mean element's abundance encountered by the CBs along
their trajectory in the ISM around a distance  where the CB Lorentz 
factor is $\gamma\approx E/A\,  m_p\, .$ The $A$-dependence follows 
from the fact that all CR nuclei, which
are accelerated by CBs, have the same universal Lorentz factor 
distribution  $dn_A/d\gamma \propto \gamma^{-p}$: The substitution
$\gamma=E/A\, m$ yields  $dn_A/dE \propto A^{p-1}\, E^{-p}.$

\noindent
More than 90\% of SN explosions take place in star formation-regions which
are enclosed in superbubbles (SB) formed by the ejecta from former SN
explosions and massive stars.  Therefor, near the GRB site, 
the elemental abundances $X_A$ are
typically that of young SNRs (which are made of SN ejecta +
progenitor ejecta prior to the SN). Further down  they become typical SB
abundances. Both, $X_A[SNR]$ and $X_A[SB]$ are not well known. 
Their metalicity enhancement is expected to progress from 
about a factor $\sim 2$ for C,N,O,Ne and $\sim 5$ 
for heavier elements such as Mg,Si,S,Ar,Ca,Fe,Ni
(e.g. Higdon et al. 2001).
Outside the SB, $X_A$ become normal ISM abundances and when the
jet enters the halo, $X_A$ become typical halo abundances. 
The decrease in metalicity 
along the CB trajectories by a factor of a few,   
induces a noticeable change 
in the spectral index of the different CR elements. 
It makes the proton spectrum slightly steeper and 
decreases slightly the steepness of the energy spectrum of the metals. 
The change in the spectral index is given roughly by 
\begin{equation}
\delta p_{_{A}}\approx  {\delta [log{X_A[E/A]}]\over
                   \delta [log(E/A)]}\, . 
\label{steep} 
\end{equation}
The change in elemental abundances along the CB
trajectory from $X_{A}\sim X_A[\odot]$ 
to $X_{A}\sim X_A[SNR]\,, $
corresponds to  energies increase from $E\sim m_{_A}$ to $E\sim
E_{max}[A]\,.$  Thus, for protons, $\delta_p\approx 0.08$ below the proton 
knee, while for iron nuclei below the iron knee,
$\delta p\approx -0.08\,.$ 
In Fig.~\ref{fig3} we compare the
predicted spectral index below the knee for different elements and their
observed values as reported by Wiebel-Sooth et al.~(1997) from their best
fits to the world CR data below the knee. The theoretical
predictions are well interpolated by  
$p \sim 2.73-0.02\, lnA\, .$  
\noindent
During their residence time in galaxies, spallation of CRs in collisions
with ISM nuclei increases significantly the abundances of the long-lived rare
elements, but change only slightly the abundances of the most abundant
elements. A detailed discussion of the effects of spallation on CR
elemental abundances, is beyond the scope of this report. 
Thus, Table I compares the observed CR elemental
abundances relative to hydrogen near TeV energy and 
the CR abundences which are predicted by Eq.~(\ref{CRabund})
for the most abundant elements He,C,N,O,Ne,Mg,Si,S,Ar,Ca,Fe,Ni. 
Table I also lists the values of $X_A[ISM]\approx X_A[\odot]$ 
(e.g. Grevesse \& Sauval 2002), which 
were used in the calculations.
Note the large enhancement in the abundances of all the elements relative 
to hydrogen. The CB model reproduces well these 
large enhancements for the abundant elements which are not changed 
much by spallation.

\noindent 
Above the proton knee, CRs become progressively poor in protons. Above the
$He$ knee, they become poor also in $He\, ,$, than also in $CNO$, etc,
until near the iron knee (the second knee) where they consist mainly of
iron and traces of heavier elements.  Spallation affect only slightly the 
value of $\langle ln A\rangle $. In Figs.~\ref{fig5},~\ref{fig6} the 
CB model prediction for
$\langle lnA\rangle$ as function of CR energy, without spallation, 
and its extracted
values from observations (see, e.g. Hoerandel 2004) are compared.  Despite 
the large spread
in the world data, the CB model clearly reproduces the observed trends. 

\section{CR acceleration above the knees} 
\noindent 
Swept in ISM particles can also be Fermi accelerated 
within the CBs by their turbulent magnetic fields to an  energy 
distribution $dN_A/dE'\propto E'^{-2}\, \Theta(E'_{min}-E')\,
\Theta(E'-E'_{max})$ within the CBs, 
where $E'_{min}=\gamma\, A\, m_p$ is the energy of the ISM 
particles entering a CB and $E'_{max}$ is the maximum energy which CRs
can be accelerated to in the CBs before their Larmor radius exceeds 
$R_{cb}$, the radius of the CBs (the lLarmor cutoff). In the lab, the  
maximal CR energy 
in the CB,  $E'_{max}\approx e\,Z\, B_{cb}\, R_{cb}\, ,$  
becomes 
\begin{equation}
E_{max}\approx max[\gamma\, E'_{max}\,(1+\beta\, \beta'\, cos\theta)]= 2\, 
\gamma\,e\, Z\, 
B_{cb}\, R_{cb}\,.
\label{maxcbe}
\end{equation}
This relation differs from the usual `Hillas relation' by the factor 
$2\,\gamma$ which is due to the relativistic motion of the CR accelerator 
-- the CB.  For a magnetic field whose pressure is equal 
to that of the scattered ISM particles (Dado et al.~2002a), 
$B_{cb}\approx \gamma\,\sqrt{2\, \pi\, n_p\, \bar{m}}\,,$
where  $n_p$ is the superbubble density, and then the maximal energy is,
\begin{equation}
E_{max}[Z]\approx 2\times 10^{20}\, Z\, \left[{\gamma_0^2\over 10^6}\right]\,
       \left[{n_p\over 10^{-3}\, cm^{-3}}\right]\,
        \left[{R_{cb}\over 10^{14}\,cm} \right]\, eV\, .
\label{maxlabe}
\end{equation}
Hence, SN jets which produce GRBs can accelerate CR nuclei to energies 
much higher than the highest energy of a cosmic ray which has ever been 
measured ($\sim 4\times 10^{20}\, eV$). 

The integrated spectrum of CRs which are Fermi accelerated within the CBs 
and escape them by diffusion can be obtained by integrating their emission 
along the trajectories of the decelerating CBs. The calculation is similar 
to the calculation of the spectrum of CR acceleration by CB elastic 
scattering, below the knee, but it is too long to be included here and 
will be described elsewhere (Dar and De R\'ujula 2005).

\section{The CR ankle}
\noindent 
In the CB model the CR ankle is the energy where the Galactic magnetic
fields can no longer isotropize the cosmic rays and delay their free
escape from the Galaxy.  This happens when the Larmor radius $R_L$ of the
CRs approaches the coherence length, $R_c\sim 1\, kpc\, $ of the turbulent
Galactic magnetic fields. Consequently, the accumulation time
of Galactic cosmic rays must satisfy,
\begin{equation}
\tau_{_{CR}}[MW]\approx 2\times 10^7\, \left[ {E\over GeV}\right]^{-r}\, y 
                \approx {R_L\over c}\, 
               \approx {R_c\over c}\approx  3\times 10^3\, y\, . 
 \label{rvalue}
\end{equation}
For $E=E_{ankle}\approx 10^{9.5}\, GeV\,,$ Eq.~(\ref{rvalue}) yields 
$r\approx 0.4\, .$ This value is between $c=1/2$ which is 
expected at low energies for a Kraichnan  spectrum of MHD 
turbulence, and $c=1/3$ which is expected at high energies for a 
Kolmogorov spectrum of MHD turbulence. The transition energy between the 
two values has not been predicted yet from the diffusion models.
A consistency check for the above interpretation of the origin of the 
ankle can be obtained as follows:  

\noindent
The accumulation time of CRs in the IGM
is roughly the age of the Universe, $\tau_{CR}[IGM]\sim
t_{_H}=1/H_0\approx  15\, Gy\,,$ where  
$H_0\sim 65 km\, s^{-1}\, Mpc^{-1}$ is the Hubble constant.
The SN rate in the Universe per comoving unit volume 
is roughly proportional to the luminosity of this volume.
The luminosity density in the local Universe is, 
$\rho_L[IGM]\approx 1.2\times 10^8\, L_\odot\, Mpc^{-3}\, .$
The optical luminosity of the Milky Way is
$L_*[MW]\approx 2.4\times 10^{10}\, L_\odot$ 
and the  volume of its CR halo 
is\footnote{This volume was estimated from a best
fit by Strong et al.~(2004) 
of the CR production rate of $\gamma$-rays
in a ``leaky box" model of
the galactic CR halo to
the diffuse gamma-ray emission of the Galaxy as measured by
EGRET (Sreekumar et al.~1998).}
$V_{CR}[MW]\approx 2.1\times 10^{68}\, cm^3$ (the volume 
of a cylinder with a diameter of $30\, kpc$ and a height of $10\, kpc\, 
.$) The extragalactic (EG) flux takes over 
the Galactic flux at $E_{ankle}\approx 10^{9.5}\, GeV\, ,$ where 
\begin{equation}
 {dF_{EG}\over dE} \sim  {\tau_{_H}\, I_{cos}\, \rho_L[IGM] \over 
                    \tau_{_{CR}}[MW]\,\rho_L[MW]}\, {dF_{MW}\over dE}
\sim {dF_{MW}\over dE}\, , 
\label{EGCR} 
\end{equation} 
i.e., the theory predicts that at $E=E_{ankle}\,, $
\begin{equation}
\tau_{_{CR}}[MW]\approx {\rho_L[IGM]\over \rho_L[MW]}\,I_{cos}\,  
\tau_{_H}\, ,
\label{EGCR1}
\end{equation} 
where the effects of cosmic expansion and stellar evolution are
included in a cosmological factor which, for the standard cosmology, has 
approximately the value
$I_{cos}\approx 2.5\pm 1.0\, ,$ as discussed in the next section. Thus,
Eq.~(\ref{EGCR1}) is well satisfied at $E=E_{ankle}\approx 10^{9.5}\, GeV$
if $c\approx 0.4\, .$  

\section{The spectrum of the extragalactic UHECRs}

The cosmic SN rate is believed to be proportional to the
cosmic star formation rate, which is given approximately by
(see, e.g. Lilly et al. 1995; Madau et al. 1996;
Perez-Gonzalez et al. 2005 and 
references therein),
$R_{SF}(z\leq 1.3)\sim R_{SF}(0)\, (1+z)^4$ and $R_{SF}(1.3\leq z\leq 
5)\approx R_{SF}(z=1.3)\, .$
In a steady state, the injection rate of CRs into the
IGM by a galaxy is equal to their galactic production rate.
Consequently, SN explosions in galaxies,
inject into the IGM the kinetic energy of
their jets at a rate $R_{SN}\, E_k $ per comoving unit volume,
where the rate of SN explosion in a
comoving unit volume, presumably, is proportional to the star
formation rate, $R_{SN}(z)=R_{SN}(0)\, R_{SF}(z)/R_{SF}(0)\,. $
Using the redshift-time relation for a standard cosmology
with $\Omega=\Omega_M+\Omega_\Lambda=1\, ,$
\begin{equation}
{dt\over dz}= {1\over H_o\, (1+z)\,
               \sqrt{\Omega_M\, (1+z)^3 +\Omega_\Lambda}}\, ,
\label{dtdz}
\end{equation}
the CR spectral density in the local IGM is given by,
\begin{equation}
{dn_{_{CR}}\over dE}= {R_{SN}(0)\over H_0}\, \int_0^\infty \,
                  {R_{SF}(z) \over R_{SF}(0)}\,
 {dN_{CR}\over dE'}\,
                  {dz\over \sqrt{\Omega_M\, (1+z)^3
+\Omega_\Lambda}}\, ,
\label{dndezIGM}
\end{equation}
where $N_{CR}$ is the number of CRs
produced by the bipolar jets in a single SN explosion, $E'=(1+z)\,E\,,$
 and consequently $dN_{CR}/dE'\propto
(1+z)^{-p}\,E^{-p}\,,$  
$\Omega_M\approx 0.27\,,$ $\Omega_\Lambda\approx 0.73\, ,$
$H_0\approx 65\, km\, Mpc^{-1}\, s^{-1}\,,$ and,
$R_{SN}(z=0)\approx 10^{-4}\, Mpc^{-3}\, y^{-1}\, ,$ is
the observed SN rate in the local Universe is\footnote{
The measured SN rate is  2.8 y$^{-1}$ (Van den Bergh \&
Tammann, 1991) in a ``fiducial volume" of 342 galaxies within the
Virgo circle whose total B-band luminosity is
$1.35\,h^{-2}\times 10^{12}\,L_\odot^B$).
For $h=0.65$ ($H_0$ in units of $100\,km\, Mpc^{-1}\, s^{-1}$) 
and the galactic luminosity
$L_*{MW}\sim 2.4 \times 10^{10}\, L_\odot $, we also obtain
$R_{SN}[MW]\approx\ 1/50\, y^{-1}$.}.
Hence, neglecting the interaction of CRs in the IGM, 
the cosmic expansion modifies the CR accumulation by 
the factor
\begin{equation}
I_{cos}=\int_0^\infty \,
                  {R_{SF}(z) \over R_{SF}(0)}\,
                  {(1+z)^{-p}\,dz\over  \sqrt{\Omega_M\, (1+z)^3
                  +\Omega_\Lambda}}\approx 2.5 \pm 1.0\, .
\label{ICOS}
\end{equation}

During their accumulation time in the IGM, the UHECRs from SN explosions
in external galaxies are completely isotropized there by the IGM magnetic
fields. Their injection spectrum, is modified in the IGM mainly by cosmic 
expansion and by photo-disintegration, pair 
production and pion production in collisions with photons of the cosmic
far infrared (FIR), microwave and radio background radiations.

In the CB model, these lead to four major predictions for CRs 
\begin{itemize}
\item{}
Most nuclei heavier than $He$ are destroyed in the IGM by 
photo-disintegration and the composition of UHECRs becomes almost 
purely protons and $He$ nuclei.   
\item{}
The spectrum of UHECR protons and $He$ nuclei must show the GZK cutoff. 
\item{}
The CR flux above the ankle should remain highly isotropic
up to the GZK cutoff. 
\item{}
The absolute normalization of the extragalactic flux 
can be calculated from the Galactic luminosity of UHECRs,
the SN rate in the local universe and the star formation rate
as function of redshift or look-back time.   
\end{itemize}

Figs. 2-7 compare the CB model predictions and the observations
of the spectrum, composition and
the depth of shower maximum of the UHECRs. The detailed
descriptions of the calculations of the spectrum, the composition and the
depth of shower maximum of the UHECRs are beyond the scope of this paper
and will be described elsewhere (Dar and De R\'ujula, 2005).
The predictions of the CB model seem to favour the results 
from the Fly's eye and HiRes experiments over those from the AGASA 
experiment,
(unless the AGASA energies are reduced by ~30\%) as can be clearly seen 
from Fig. 2.

\section{The Galactic CR Luminosity}
In a steady state, the escape rate of Galactic CRs into the IGM 
is equal to their production rate. In the CB model, most of the kinetic 
energy of SN jets is converted to CR energy in the Galaxy. 
In the CB model, this energy  was estimated {Dar \& De R\'ujula~2000,2004)
to be $E_k\sim 2\times 10^{51} erg$ per SN. Thus, a Galactic SN 
rate, 
$R_{SN}\sim 1/50\, yr^{-1}\,,$ generates a Galactic CR luminosity of 
\begin{equation}
   L_{cr}[MW]=R_{SN}\, E_k\approx 1.3\times 10^{42}\, erg\, s^{-1}. 
\label{LCR}
\end{equation}
\noindent
This value is consistent with a more direct estimate of $L_{CR}[MW]\, :$
As a result of the steep energy spectrum of galactic CRs, the bulk of the
CR energy is carried by nuclei with an average energy of a few GeV.
The most accurate  measurements of their flux, $dI/dE$, near Earth and
during solar minimum (minimum solar modulation) are those of
AMS (Alcaraz et al.~2000a,b) and BESS (Haino et al.~2004).
Under the assumption of constant CR density inside the Galactic CR halo, 
their measurements yield a  CR Luminosity \footnote{At  energies
below $\sim\! 10$ GeV,
$dn/dE$ is affected by the Earth's and solar wind magnetic fields. The 
uncertainties in the
energy-weighed  integral of Eq.~(\ref{LECR}) are much smaller.}:
\begin{equation}
   L_{cr}[MW]\approx  
={4\,\pi V_{CR}[MW]\over c}\,\int {\ dn_{CR}[MW]\over dE}\, {E\, dE\over 
\tau_{_{CR}}[E]}\approx 1.1\times 10^{42}\, erg\, s^{-1}. 
\label{LECR}
\end{equation}
Traditional estimates which are based on an assumed $\sim 10\%$ conversion 
of the kinetic energy of the non-relativistic ejecta in SN explosions  
to CR energy through collisionless shock acceleration, yield much smaller 
CR luminosities, $L_{cr}< 10^{41}\, erg\, s^{-1}\,.$ They are adjusted to 
fit estimates from CR diffusion models which ignore the direct deposition 
of CRs by SN jets in the Galactic halo and the in the IGM.

\section{Conclusions} 
\noindent 

The CB model of GRBs was used here to demonstrate that the bipolar jets 
from SN explosions can be the main source of cosmic rays at all energies.  
The model correctly predicts, within the experimental uncertainties, the 
observed all-particle CR flux, its energy spectrum and its elemental 
composition at all energies.  In particular, the model predicts that the 
CRs above the ankle are extragalactic in origin, are mostly protons $+He$ 
nuclei, are highly isotropic and their spectrum exhibit the GZK suppression 
due to their interaction with the microwave and radio background 
radiations. It appears like the 93 years old puzzle of the origin of 
galactic CRs has been solved. However, more precise CR data on their 
energy spectrum and composition, in particular near the CR knees and well 
above the GZK cutoff, as well as astrophysical data on the elemental 
abundances in SNRs and SBs are needed before a firm conclusion can be 
made.

\section{Acknowledgments} 
The author would like to thank Shlomo Dado,
and Alvaro De R\'ujula  for a long term collaboration  
in the development of the CB model for GRBs and CRs.
The support of the Asher Space Research Fund  
is gratefully acknowledged. This research was not supported 
by the Israel Science Foundation for basic research.

\begin{deluxetable}{llcccc}
\tablewidth{0pt}
\tablecaption{Solar and CR abundances at 1 TeV relative 
to hydrogen.}

\tablehead{\colhead{ } &
\colhead{Z} &\colhead{$X_\odot$} & \colhead{$X_{_{\rm CR}}$}&
\colhead{$A^{1.73}\, X_{_{\rm SB}}$}&} 
\startdata

H  & 1 &  1                   & 1   &  1  \\

He & 2 & $7.5\times 10^{-2}$ & $6.5\times 10^{-1}$
& $8.2\times 10^{-1}$\\

\tableline

C  & 6 & $3.3\times 10^{-4}$ & $9.2\times 10^{-2}$
& $4.9\times 10^{-2}$\\

N  & 7 & $8.3\times 10^{-5}$ & $2.0\times 10^{-2}$
& $1.6\times 10^{-2}$\\

O  & 8 & $6.7\times 10^{-4}$ & $1.4\times 10^{-1}$
& $1.6\times 10^{-1}$\\

Ne & 10& $1.2\times 10^{-4}$ & $3.8\times 10^{-2}$
& $ 4.3\times 10^{-2}$\\

\tableline

Mg & 12& $3.8\times 10^{-5}$ & $6.7\times 10^{-2}$
& $4.8\times 10^{-2}$\\

Si & 14& $3.5\times 10^{-5}$ & $6.9\times 10^{-2}$
& $5.6\times 10^{-2}$\\

S  & 16& $2.1\times 10^{-5}$ & $2.0\times 10^{-2}$
& $4.2\times 10^{-2}$\\

Ar & 18& $2.5\times 10^{-6}$ & $7.3\times 10^{-3}$
& $7.4\times 10^{-3}$\\

\tableline

Ca & 20& $2.3\times 10^{-6}$ & $1.3\times 10^{-2}$
& $6.8 \times 10^{-3}$\\

Fe & 26& $3.2\times 10^{-5}$ & $1.6\times 10^{-1}$
& $1.7\times 10^{-1}$\\

Ni & 28& $1.8\times 10^{-6}$ & $8.6\times 10^{-3}$
& $1.0\times 10^{-2}$\\

\tableline
\enddata
\end{deluxetable}

\newpage
\begin{figure}
\vspace*{13pt}
\leftline{\hfill\vbox{\hrule width 5cm height0.001pt}\hfill}
        \mbox{\epsfig{figure=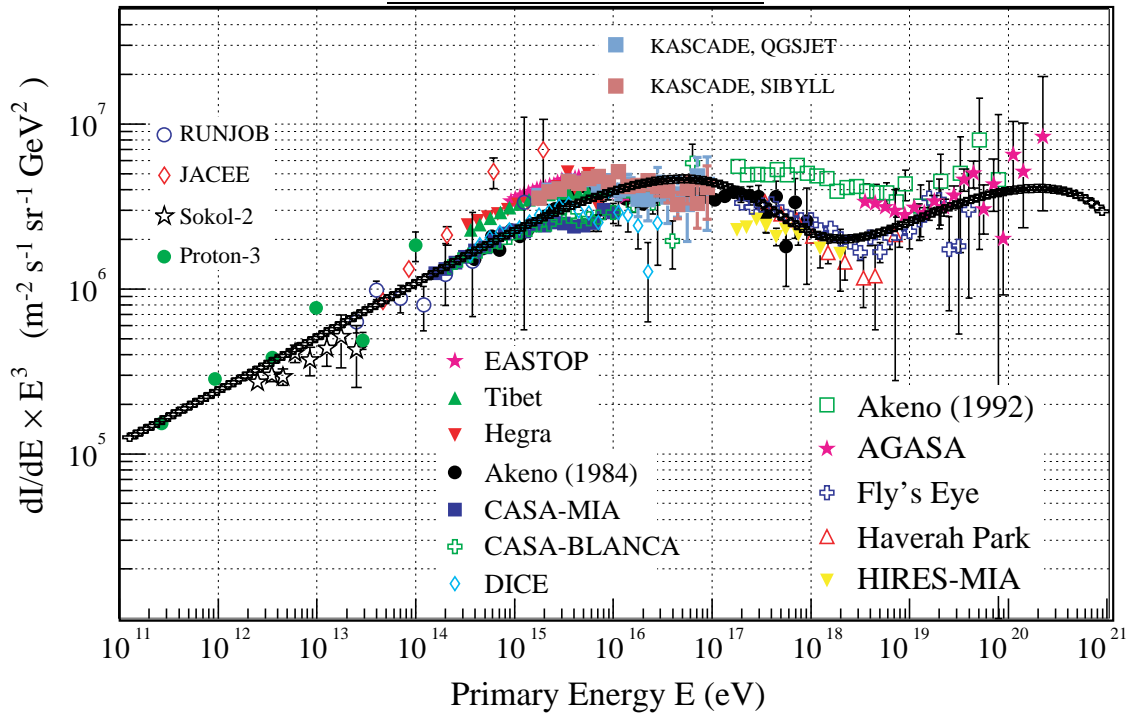,width=15.0cm}}
\vspace*{0.2truein}            
\leftline{\hfill\vbox{\hrule width 5cm height0.001pt}\hfill}
\caption{Comparison between the CB model prediction  (crosses) 
for  the all-particle CR spectrum before the inclusion 
of CR interactions in the IGM
and the world data as compiled by Ulrich (Kampert et al. 2004).
The effects of cosmic expansion and star formation rate were included.
The high energy decline is due to the Larmor cutoff. The theoretical
predictions and the  observations have been multiplied in both figures 
by the third power of the energy to emphasize significant deviations from 
a single  power-law 
decline over thirty orders of magnitude.}
\label{fig1}
\end{figure}

\newpage
\begin{figure}
\vspace*{13pt}
\leftline{\hfill\vbox{\hrule width 5cm height0.001pt}\hfill}
        \mbox{\epsfig{figure=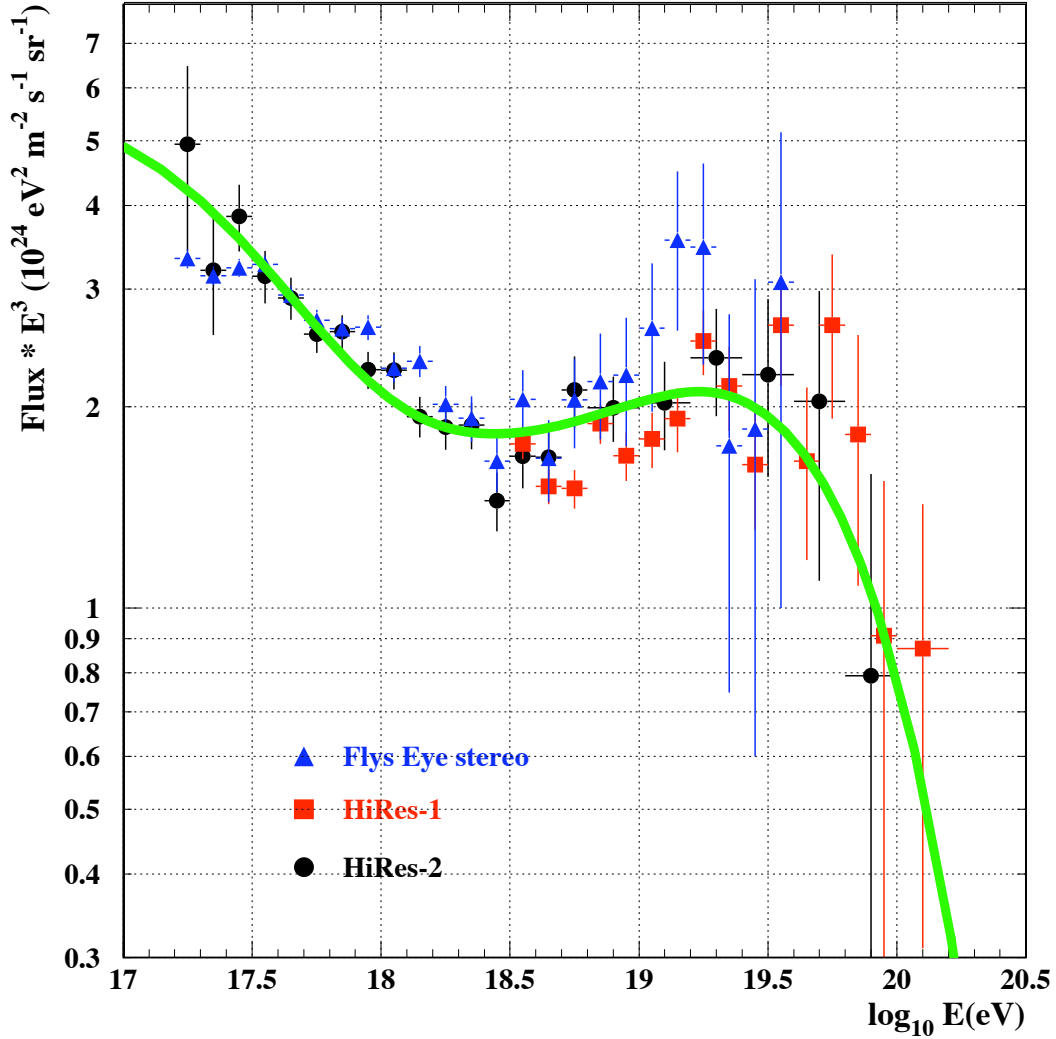,width=15.0cm}}
\vspace*{0.2truein}
\leftline{\hfill\vbox{\hrule width 5cm height0.001pt}\hfill}
\caption{Comparison between the HiRes data (Zech 2004)
for the all-particle CR spectrum above the knee 
and the CB model prediction 
after the inclusion of the effects of pair production,  
photo-disintegration and pion photo-production 
in collisions of CR nuclei 
with the far infrared, microwave and 
radio background photons in the IGM. The theoretical
predictions and the  observations have been multiplied in both figures by
the third power of the energy
to emphasize significant deviations from a single power-law decline.}
\label{fig2}
\end{figure}

\newpage
\begin{figure}
\vspace*{13pt}
\leftline{\hfill\vbox{\hrule width 5cm height0.001pt}\hfill}
        \mbox{\epsfig{figure=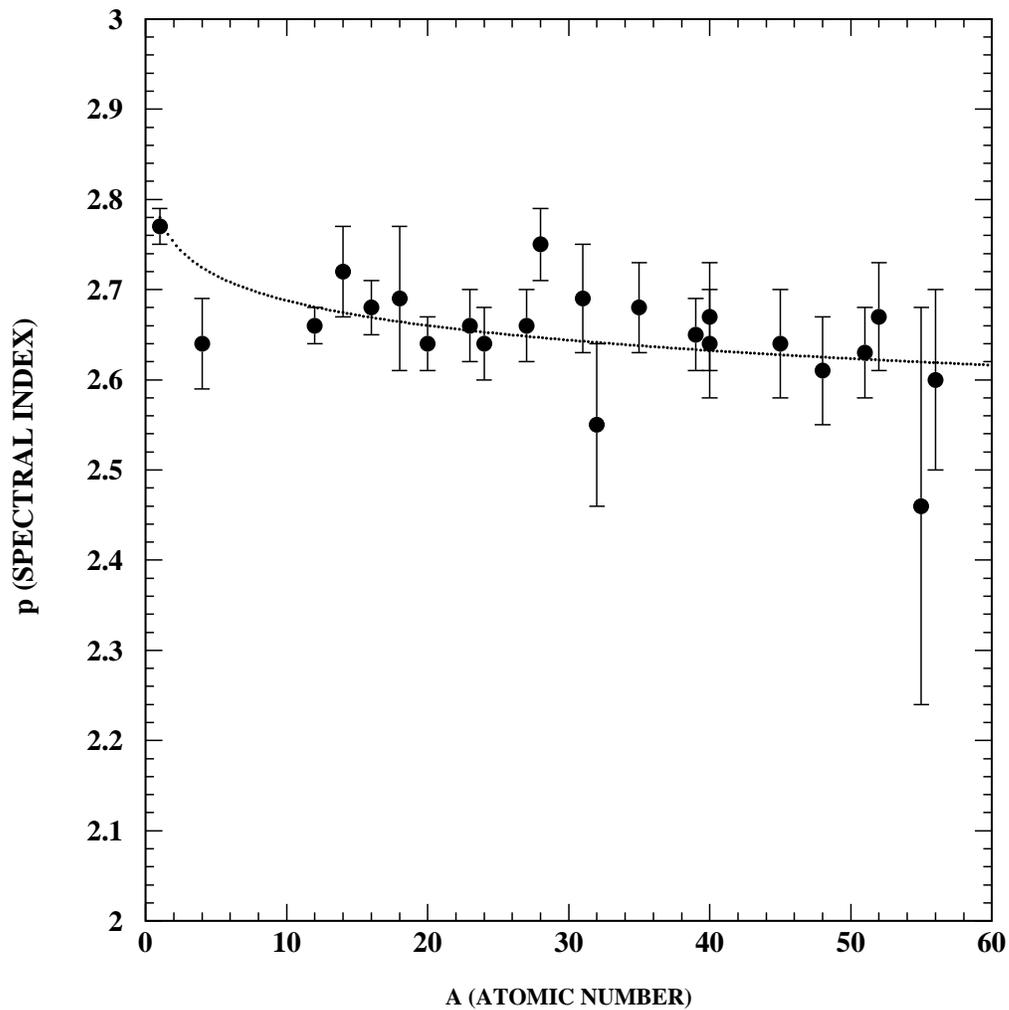,width=15.0cm}}
\vspace*{0.4truein}
\leftline{\hfill\vbox{\hrule width 5cm height0.001pt}\hfill}
\caption{Comparison between the CB model prediction for the
A-dependence of the spectral index of CRs  for energies below
the knee, Eq. 18, and  their values obtained by Wiebel-Sooth et al.~
(1997) from their best fits to the world CR data.}
\label{fig3}
\end{figure}

\newpage
\begin{figure}
\vspace*{13pt}
\leftline{\hfill\vbox{\hrule width 5cm height0.001pt}\hfill}
        \mbox{\epsfig{figure=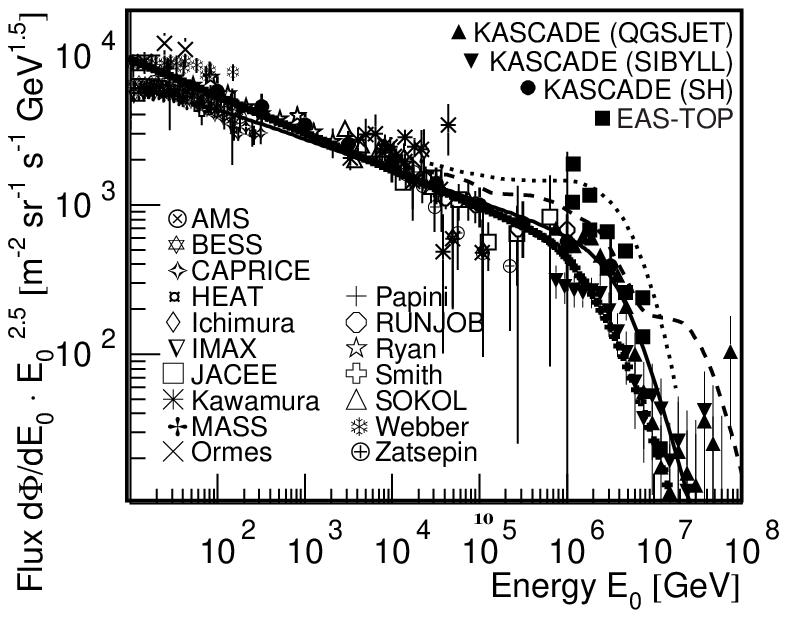,width=7.0cm}}
\vspace*{0.1truein}
\leftline{\hfill\vbox{\hrule width 5cm height0.001pt}\hfill}
        \mbox{\epsfig{figure=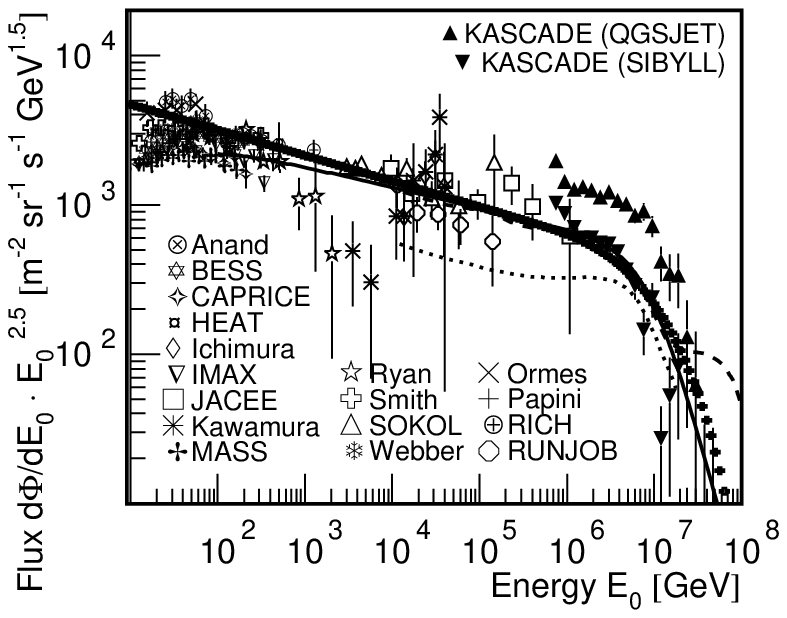,width=7.0cm}}
\vspace*{0.1truein}
\leftline{\hfill\vbox{\hrule width 5cm height0.001pt}\hfill}
        \mbox{\epsfig{figure=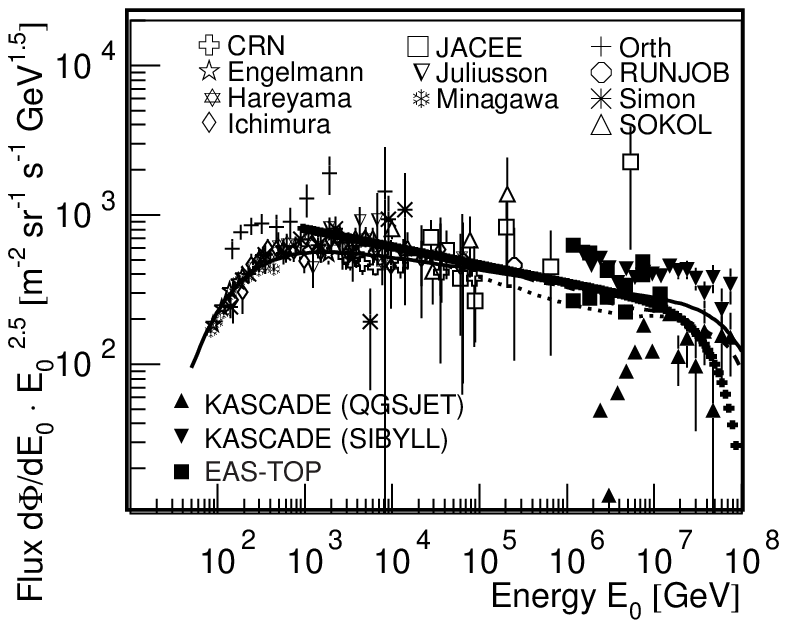,width=7.0cm}}
\vspace*{0.1truein}
\leftline{\hfill\vbox{\hrule width 5cm height0.001pt}\hfill}
\caption{Comparison between the CB model prediction (thick line)  for 
the spectrum of CRs protons (top), He (middle) and Fe nuclei (bottom)
near the knee and their inferred spectra from the KASKADE 
measurements (Hoerandel 2004).}    
\label{fig4}
\end{figure}

\newpage
\begin{figure}
\vspace*{13pt}
\leftline{\hfill\vbox{\hrule width 5cm height0.001pt}\hfill}
        \mbox{\epsfig{figure=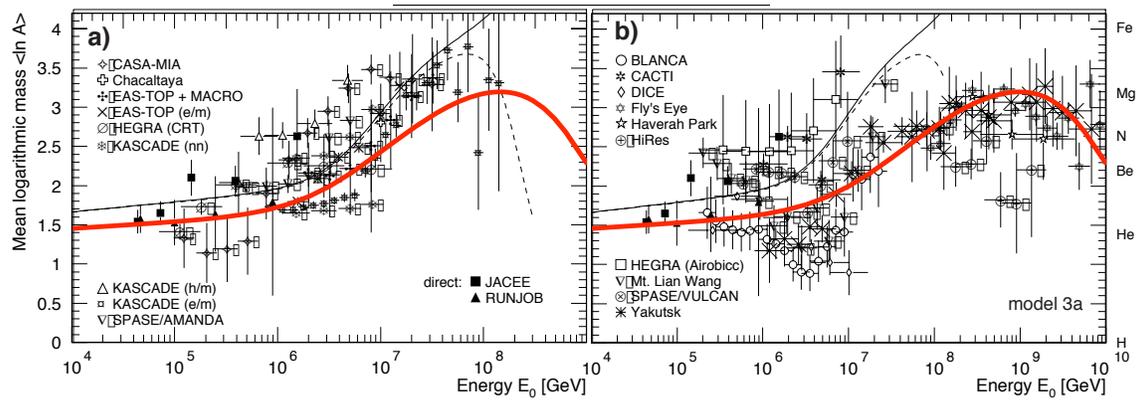,width=15.0cm}}
\vspace*{0.4truein}
\leftline{\hfill\vbox{\hrule width 5cm height0.001pt}\hfill}
\caption{Comparison between the CB model prediction for lnA as function 
of CR energy and the world data as compiled by Hoerandel (2004).} 
\label{fig5}
\end{figure}

\end{document}